\documentstyle[preprint,prl,aps,epsfig,here]{revtex}
\begin{document}
\voffset=-5pt
\hoffset=-10pt
\def\sn2{$\sin^22\theta$}
\def\dm2{$\Delta m^2$}
\def\ch2{$\chi^2$}
\def\ltap{\ \raisebox{-.4ex}{\rlap{$\sim$}} \raisebox{.4ex}{$<$}\ }
\def\gtap{\ \raisebox{-.4ex}{\rlap{$\sim$}} \raisebox{.4ex}{$>$}\ }
 \draft
\begin{titlepage}
\preprint{\vbox{\baselineskip 10pt{
\hbox{hep -- ph/9708308}
\hbox{August 1997}
\hbox{}}}}
\vskip 2truecm
\title{ \bf Mixed MSW and Vacuum Solutions of Solar Neutrino 
    Problem \footnote{Talk given at the $Fourth~ International~
Solar~Neutrino~Conference$ Heidelberg, Germany, April 8-14, 1997. }}

\author{Qiu-Yu Liu}

\address{ Scuola Internazionale Superiore di Studi Avanzati, I-34013
Trieste, Italy}
\maketitle
\begin{abstract}
\thispagestyle{empty}
\begin{minipage}{5in}
\thispagestyle{empty}
\baselineskip 16pt
Assuming three flavour neutrino mixing takes place in vacuum, 
we investigate the possibility
that the solar $\nu_e$ take part in MSW transitions in the Sun
due to $\Delta m^2_{31} \sim (10^{-7} - 10^{-4})~eV^2$, followed by long
wave length vacuum oscillations on the way to the Earth, triggered by
$\Delta m^2_{21}$ (or $\Delta m^2_{32}$) $\sim (10^{-12} - 10^{-10})~eV^2$.
The solar $\nu_e$ survival probability is shown to be described in this 
case by a simple analytic expression. 
 New ranges of neutrino 
 parameters which allow to fit the solar neutrino data have been
 found. The best fit characterized by the
 minimum $\chi^2$ is extremely good .  
 This hybrid (MSW+vacuum oscillations) solution of the solar neutrino problem 
 leads to peculiar distortions of energy spectrum of the boron neutrinos  
 which can be observed by the SuperKamiokande and SNO experiments.    
 Other flavour scheme (e.g. 2 active $\nu$s + 1 sterile $\nu$) 
 can provide MSW+vacuum solution also.
\end{minipage} 
\end{abstract}
\end{titlepage}

\newpage

\hsize 17.0truecm
\textheight = 714.2857143pt
 \normalsize
\baselineskip  21pt
\def\dm{$\Delta m^2$\hskip 0.1cm }
\def\dmsqua{$\Delta m^2$\hskip 0.1cm}
\def\sn{$\sin^2 2\theta$\hskip 0.1cm }
\def\snf{$\sin^2 2\theta$}
\def\trna{$\nu_e \rightarrow \nu_a$}
\def\trnm{$\nu_e \rightarrow \nu_{\mu}$}
\def\trns{$\nu_e \leftrightarrow \nu_s$}
\def\trnat{$\nu_e \leftrightarrow \nu_a$}
\def\trnmt{$\nu_e \leftrightarrow \nu_{\mu}$}
\def\trne{$\nu_e \rightarrow \nu_e$}
\def\trnst{$\nu_e \leftrightarrow \nu_s$}
\def\nue{$\nu_e$\hskip 0.1cm}
\def\numu{$\nu_{\mu}$\hskip 0.1cm}
\def\nutau{$\nu_{\tau}$\hskip 0.1cm}
    
\font\eightrm=cmr8
\def\aprle{\buildrel < \over {_{\sim}}}
\def\aprge{\buildrel > \over {_{\sim}}}
\renewcommand{\thefootnote}{\arabic{footnote}}
\setcounter{footnote}{0}

\leftline{\bf 1. Introduction}
\vskip 0.3cm

\indent Solar neutrino experiments
\cite{CHLOR,KAM,GALLEX,SAGE,SNP,SNP1}
indicate an existence of solar neutrino problem. Scientists have tried
several different solutions, for example, astrophysical solutions ( 
which implys the puzzle comes from our lack of the knowledge of
nucleon inside the Sun. But this solution doesn't fit data well );
spin and spin-flavour oscillations
(caused by magnetic field) \cite{AKH} and neutrino decay (can not fit 
experimental data well ). In particular, there are two 
solutions which give good
fit of data: First based on the old idea of Pontecorvo \cite{Pont1} that 
the solar $\nu_e$ take part in vacuum oscillations
when they travel from the Sun to the Earth. The second based on the more 
recent hypothesis \cite{MS,LW} of the solar $\nu_e$ undergoing
matter-enhanced (MSW) transitions into neutrinos of a different type
when they propagate from the central part to the surface of the Sun. 

     Both of MSW and vacuum oscillation mechanisms may simultaneously happen
in our nature. That is, MSW effect
 take place
when the neutrino travel from the centre of the Sun to the surface of the
Sun, long wavelength effect occur afterwards during the travel from
the surface of the Sun to the surface of the Earth. Thus, 
$\nu_{\odot}$-problem can be solved by a hybrid solution \cite{QLSP,AS91}.

    This requires three flavour neutrinos scheme which is also a motivation for
the mixed solution. There are two independent $\Delta m^2$ now. For
solving the $\nu_{\odot}$-neutrino problem, there can be sets of magnitudes of
two $\Delta m^2$ \cite{AS91}. The interesting one is that a $\Delta
m^2$ lies in the MSW mass interval and the other lies in the long wavelength
interval, corresponding to 
$$10^{-6}eV^2 \ltap (\Delta m^2)_{\rm big}
      \ltap 2\times 10^{-4}eV^2, ~\eqno(1a)$$
$$10^{-12}eV^2 \ltap (\Delta m^2)_{\rm small}
      \ltap 5\times 10^{-10}eV^2 ~\eqno(1b)$$
       
   Detailed studies of the vacuum oscillation or MSW
transition solution of the solar neutrino problem
under the more natural assumption of three
flavour neutrino mixing are still lacking \cite{PAN91}, partly because
of the relatively large number of parameters involved.
In this hybrid solution, two possible mass spectra are shown
in FIG. 1. $~$ In first case, one $m$ is heavy but the other two are
light. In the second case, two neutrino masses are heavy and the other
one is light. Both two cases are

\begin{figure}[H]
\hskip 3.5cm
\mbox{\epsfig{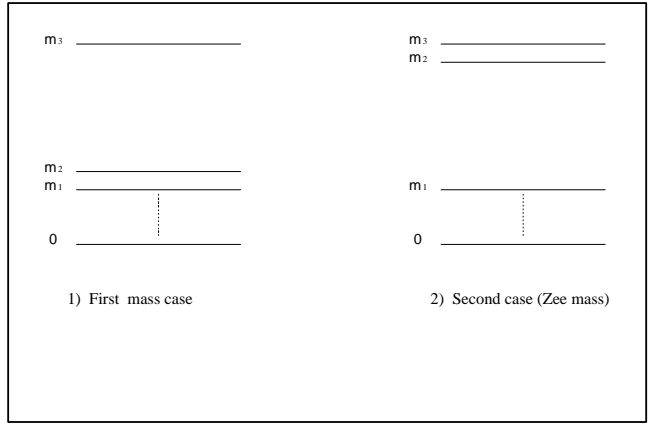}}
\caption[]{ Different mass spectra, two light plus one relative heavy
 neutrinos 1) and one light with two heavy neutrinos 2),
 which can have mixed solutions to solar neutrino problem. } 
\end{figure}

\noindent independent from the absolute value of $m$ but the difference.\\
\indent All the different 
patterns of neutrino masses mentioned above
can arise in gauge theories of electroweak interactions with 
massive neutrinos, and in particular, in GUT theories \cite{BQYAS}. 
\vskip 0.4cm
\leftline{\bf 2. The Solar \nue Survival Probability}
\vskip 0.3cm
  For the first mass case, the averaged survival probability (FIG. 2) can be written
\footnote{ When $\theta_{13}$ is very small, an approximate formula is given
in \cite{AS91}.}:
$$\bar{P}(\nu_e \rightarrow \nu_e;t_E,t_0) =
 \bar{P}^{(31)}_{2MSW}(\Delta m^2_{31},\theta_{13})-V ,$$
here $V$ is the loss of \nue caused by vacuum oscillation 
during the way from surface of the
Sun to the Earth. It can be written in several forms:
$${\displaystyle V={2|U_{e1}|^2|U_{e2}|^2 \over {1-|U_{e3}|^2}}(1-cos2\pi {R \over
{L^{v}_{21}}})[{1 \over 2}+({1 \over 2}-P_{jump}^{(31)})
cos2\theta_{31}^m]}~~~~~~~~~~~~~~~~~~~~~~~~~~~~~~~~~~~~~~~~~~~~~~~~~~$$
$$~~~~{\displaystyle ={{1 - |U_{e3}|^2}\over {1 - 2|U_{e3}|^2}}~[1 -
  P^{(21)}_{2VO}(\Delta m^2_{21},\theta_{12})]~
  [|U_{e3}|^2 - \bar{P}^{(31)}_{2MSW}(\Delta m^2_{31},\theta_{13})]~~~~~~~~~~~~~~~~~~~~~~~~~~~}$$    
 $$~~~~~~~{\displaystyle ={1\over {2}}~\sin^22\theta_{12}~(1 - \cos2\pi {R \over
{L^{v}_{21}}})~
\cos^2\theta_{13} [{1\over {2}} + ({1\over {2}} - P_{jump}^{(31)})
\cos2\theta_{13}^{m}(t_0)].~~~~~~~~~~~~~~~~~~~~~~~~~~~~~~}$$
Some comments are given below for these formulas: \\
  $\bullet$ the term $\bar{P}(\Delta m^2_{31},\theta_{13})$ is 2-neutrinos
 MSW transition $\nu_e
\Leftrightarrow \nu'$ survival probability with that
$\nu'$ is the mixed state of $\nu_{\mu}$ and $\nu_{\tau}$:
$$\nu'={U_{\mu 3} \over \sqrt{U_{\mu 3}^2+U_{\tau 3}^2}}\nu_{\mu}+e^{-i{\varphi
\over 2}}{U_{\tau 3} \over \sqrt{U_{\mu 3}^2+U_{\tau 3}^2}}\nu_{\tau},$$
where $\varphi$ is the phase of $U_{e3}$ \\
  $\bullet$ $P^{(21)}_{2VO}$ is the 2-neutrinos vacuum oscillation 
probability
with parameters $\Delta m^2_{21}$ and $\theta_{12}$.\\
  $\bullet$ The mixing angles are defined as
$$\sin^2\theta_{13}=|U_{e3}|^2;$$
$$\sin^22\theta_{12}={4|U_{e1}|^2|U_{e2}|^2 \over {(1-|U_{e3}|^2})^2}$$
which coincide with the definition below
$$\nu_e=\cos\theta_{13}\nu^{(12)}+\sin\theta_{13}\nu_3;$$
$$\nu^{(12)}=\cos\theta_{12}\nu_1+\sin\theta_{12}\nu_2,$$
where $\nu_1$, $\nu_2$ and $\nu_3$ are three neutrino mass eigenstates.  
\begin{figure}[H]
\hglue-1cm
 \epsfig{figure=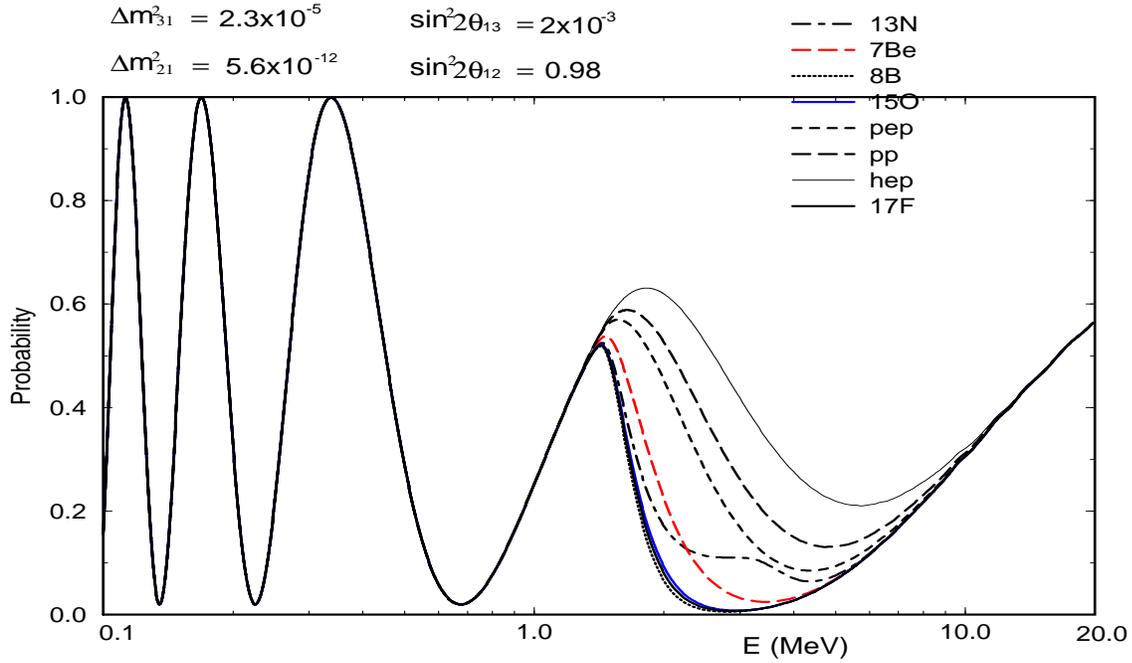,width=10cm,height=15cm,angle=-90}
\caption[]{The survival probabilities 
$\bar{P}(\nu_{e} \rightarrow \nu_{e})~$ of all types of solar
neutrinos for the first mass case. Differences is because of different
production area inside the Sun. }
\label{Psall}
\end{figure}

   Let us consider the properties of the survival probability $\bar{P}$.
For module of term $V$, its coefficient is
$$F_1\equiv {2|U_{e1}|^2|U_{e2}|^2 \over {1-|U_{e3}|^2}}={1\over
{2}}~\sin^22\theta_{12}\cos^2\theta_{13},$$
here $\cos^2\theta_{13}$ is in interval [0.5, 1], so $0\leq
F_1\leq {\displaystyle {1 \over 2}}$, i.e., the module of $V$ can vary 
from 0 to 1. The upper bound of $\bar{P}$ is $\bar{P}_{2MSW}$. Thus for 
same MSW parameters, the suppression of mixed scheme is stronger than
that of 2-$\nu$ MSW. In general, in order to recover the same theoretical 
predictions,
$\Delta m^2_{31}$ should shift to a bigger value than 2-$\nu$ MSW's 
\cite{MSW1,KLP}. 
So for the same $\sin^22\theta_{13}$, the new
allowed solution regions in MSW parameter plane must be always above the
2-$\nu$ MSW's (FIG. 4).\\
  \indent At the bottom of the MSW suppression pit, (i.e., 
$P^{jump}_{(31)}=0$ and $\cos2\theta_{13}^{m}=-1$), term $V$ goes to zero and
there is no vacuum oscillation. 
 Under the above conditions the $\nu_e$ state in matter at the
point of 
$\nu_e$ production (in the Sun) essentially coincides with the heaviest
of the three neutrino matter-eigenstates, which continuously evolves
(as the neutrino propagates towards the surface of the Sun)
into the mass (as well as energy) eigenstate $|\nu_3>$
at the surface of the Sun. As a consequence,
vacuum oscillations do not take place between the Sun and the Earth, and
$\bar{P}(\nu_e \rightarrow \nu_e ; t_E,t_0)$ coincides with the
probability
to find $\nu_e$ in the state $|\nu_3>$.

 There are two limits. When  $\sin^22\theta_{12} \rightarrow 0$, we get  
 $\bar{P} \rightarrow \bar{P}_{2MSW}$; when $\sin^22\theta_{13}
 \rightarrow 0$ and $\Delta m^2_{31} \rightarrow \infty$, then 
$\bar{P} \rightarrow \bar{P}_{2VO}$.
\vskip 0.6truecm
For the second mass spectrum case \cite{ASTZee}, after exchanging the indices $1 
\leftrightarrow 3$, the term $V$ in probability still has some difference from
that in case 1, 
$$V={1\over {2}}~\sin^22\theta_{23}~(1 - \cos2\pi {R \over
{L^{v}_{32}}})~
\sin^2\theta'_{13} [{1\over {2}} - ({1\over {2}} - P_{jump}^{(13)})
\cos2\theta_{13}^{'m}(t_0)],~~~~~~~~$$ 
where $\sin^2\theta_{13}'$ is less than ${\displaystyle	{1 \over 2}}$.
  Zee model (radiative) mass matrix corresponds to this case \cite{Zee},
it reduces the four free parameters to three free parameters.
  Here module of $V$ varys from 0 to ${\displaystyle {1 \over 2}}$. 
Since the coefficient 
 $F_2={\displaystyle {1\over 2}}~\sin^22\theta_{32}\sin^2\theta'_{13}$ in $V$ is from 0 to 
${\displaystyle {1 \over 4}}$, thus the vacuum oscillation is 
suppressed by factor 2 than in the first case. Furthermore, for small 
$\sin^2\theta'_{13}$, $F_2$ is small. So we shouldn't expect new solutions in 
small MSW parameter region.  
The maximal vacuum oscillation take place at the bottom of the
MSW pit. In the two ``asymptotic'' regions ( highest energy and lowest energy
regions), $|V|~=~\sin^4\theta'_{13}~\leq~ {\displaystyle {1 \over 4}}$ 
(see FIG. 3).

\begin{figure}[H]
\hglue-1cm
 \epsfig{figure=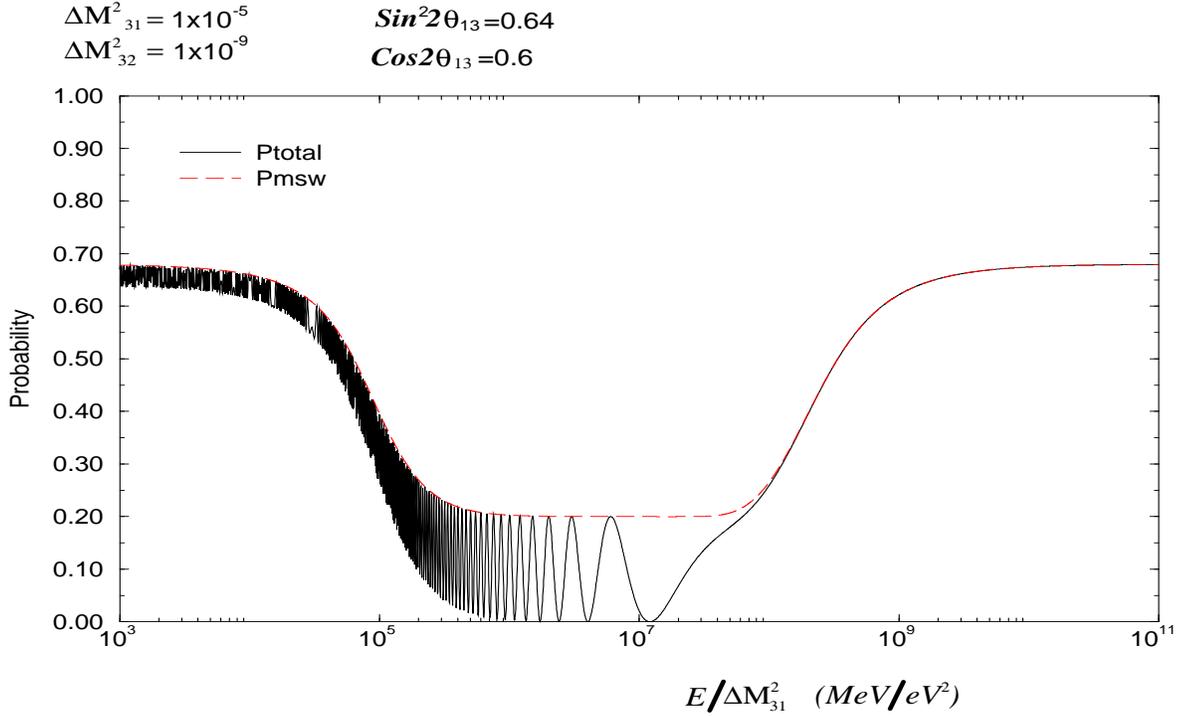,width=10cm,height=15cm,angle=-90}
\caption[]{The survival probability  
$\bar{P}(\nu_{e} \rightarrow \nu_{e})~$ (solid line) for  
Zee mass case. Dashed line is pure MSW probability. }
\label{PsZee}
\end{figure}

\leftline{\bf 3. Hybrid MSW Transition + Vacuum Oscillation Solutions} 
\hskip 0.3truecm {\bf of the Solar Neutrino Problem }
\vskip 0.3cm
\indent We have analyzed the published data from the
four solar neutrino experiments \cite{CHLOR,KAM,GALLEX,SAGE} searching for solutions 
of the solar neutrino 
problem of the hybrid MSW transitions + vacuum oscillations type. Only the
 first case of mass spectra with $\Delta m^2_{31}$ and $\Delta m^2_{21}$ 
having values
in the intervals (1a) and (1b) was studied. 

 We utilized the predictions of the solar model of Bahcall and
Pinsonneault in 1995 
with heavy element diffusion \cite{BP95} for the $pp$, $pep$, etc. neutrino fluxes 
in this study. The estimated uncertainties in the theoretical
predictions for the indicated fluxes \cite{BP95} were not taken into account.
The solutions fit to data are called A,B,C,D,E,F as below. 

 {\bf Solution A and D.} 
   
For solution A, 
The minimum $\chi^2$ ($\sim$ 0.1) value is reached at around point
($\Delta m^2_{21}$, $\sin^22\theta_{12}$, $\Delta m^2_{31}$, 
$\sin^22\theta_{13}$) $\cong$ ($5.6\times 10^{-12}~eV^2$, 0.98, $4.2\times 10^{-5}~eV^2$, $10^{-3}$). 

   For $E \geq 5~MeV$ and $\Delta m^2_{21}\leq 8.0\times 10^{-12}~eV^2$
most of the $^{8}$B neutrinos undergo only MSW transitions
\footnote{Obviously, if, for instance,  
 $\Delta m^2_{21} = 4.0\times 10^{-12}~eV^2$, the same result will be valid
for the neutrinos with $E \geq 2.5~MeV$.}. 
The MSW transitions of the $^{8}$B neutrinos having this energy
are adiabatic for values of 
$\Delta m^2_{31} \cong (1.1 - 1.3) \times 10^{-4}~eV^2$
and $\sin^22\theta_{13}\gtap (3.0 - 4.0)\times 10^{-3}$ 
from the ``horizontal'' region of the solution (see Fig. 4a).
They are nonadiabatic for values of 
$\Delta m^2_{31}$ and $\sin^22\theta_{13}$ 
from the remaining part of the allowed region.

\begin{figure}[H]
\hglue 1.5cm
\mbox{\epsfig{figure=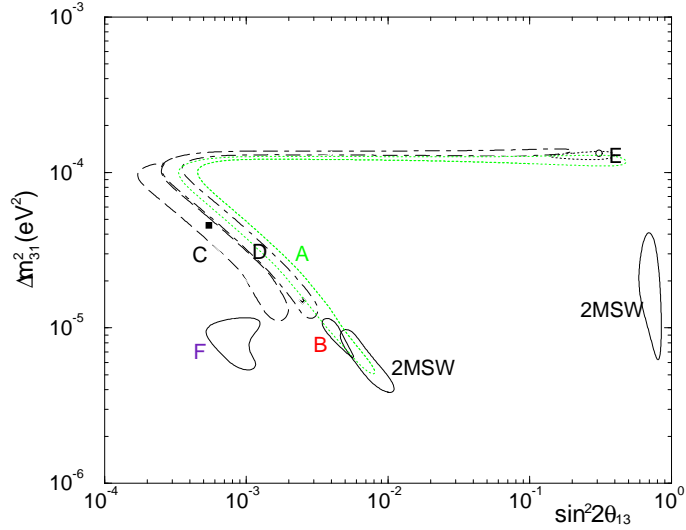,width=10cm}(a)}
\vglue0.2cm
\hglue 1.5cm
\mbox{\epsfig{figure=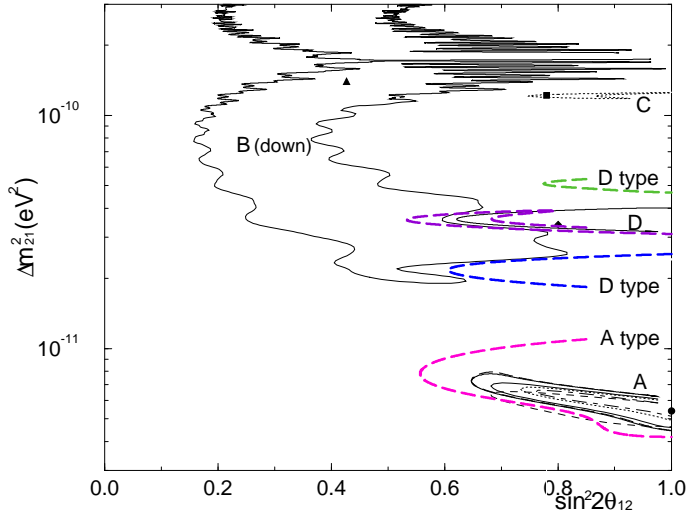,width=10cm}(b)}
\vglue0.5cm
\caption[]{ Fits to the solar neutrino data 
: (a) in MSW parameters plain, and (b) in vacuum's plain. For
solutions B, F in (a), SuperK result is included. } 
\end{figure}

 In contrast to the main fraction of $^{8}$B neutrinos, 
the $pp$ and the high energy line of $^{7}$Be neutrinos 
do not undergo resonant MSW transitions but 
take part in vacuum oscillations between the Sun and the Earth. 
Actually, the $^{7}$Be neutrino energy 
of $0.862~MeV$ is in the region of the first minimum of 
$P^{(21)}_{2VO}$ as $E$ decreases from the ``asymptotic'' values at which 
$P^{(21)}_{2VO} \cong 1$, while the interval of energies of the $pp$ neutrinos,
relevant for the current Ga--Ge and the presently discussed 
future solar neutrino experiments (HELLAZ, HERON), 
$0.22~MeV \ltap E \leq 0.41~MeV$, is in the region of the first 
maximum of $P^{(21)}_{2VO}$ as $E$ decreases further. 

   
  Let us note that the solution A region in the 
$\Delta m^2_{21} - \sin^2\theta_{12}$ plane of the vacuum oscillation
parameters is quite similar to the region
of the ``low'' $^{8}$B \nue flux 
$2\nu$ vacuum oscillation solution found in ref. \cite{KP1}. The latter is 
possible for values of the $^{8}$B neutrino flux which are lower by a factor
of 0.35 to 0.43 (of 0.30 to 0.37) than the flux predicted in \cite{BP92}
(in \cite{BP95} and used in the present study). Note, however, that the
$\chi^2_{min}$ for the indicated purely vacuum oscillation solution,
$\chi^2_{min} = 4.4~ (2~d.f.)$ \cite{KP1}, is
considerably larger than the value of $\chi^2_{min}$ for solution A. 
The two solutions differ drastically in the way the $^{8}$B 
neutrino flux is affected by the transitions and/or the oscillations.

Let us mention only here that in the case of solution
A: i) the spectrum of $^{8}$B neutrinos will be strongly 
deformed, 
ii) the magnitude of the day-night asymmetry  
in the signals in the indicated detectors 
can be very different from that predicted 
in the case of the $2\nu$ MSW solution (see,e.g., \cite{LMP,DN}), 
and iii) the seasonal variation of
the $^{8}$B $\nu_e$ flux \cite{SMBP78,KP2} practically coincides with the 
standard (geometrical)
one of 6.68\%.\\  
\indent Solution D are similar to A but $\Delta m^2_{21}$ is bigger than
 which in A.   
 It has as a $\theta_{13}\rightarrow 0$ limit the second
``low'' $^{8}$B $\nu_e$ flux $2\nu$ vacuum oscillation solution discussed in
ref. \cite{KP1}: the regions of values of the vacuum oscillation parameters
of the two solutions practically coincide. This $2\nu$ solution  
was found to be possible for values of the initial   
$^{8}$B $\nu_e$ flux which are by a factor $\sim (0.45 - 0.65)$ 
($\sim (0.39 - 0.56)$) smaller than
the flux predicted in ref. \cite{BP92} (in ref. \cite{BP95}). 
The MSW + VO effects and correspondingly the purely VO effects on
the $^{7}$Be and/or $^{8}$B neutrino fluxes in the cases of 
the two solutions are also very different.

 {\bf Solution B.} 

This solution (see FIG. 4) 
 can be regarded as
an ``improved'' MSW transitions + vacuum oscillations version of the
purely $2\nu$ MSW nonadiabatic solution.    

\indent For $\Delta m^2_{21}$ from the domain 
$\sim (0.5 - 1.0)\times 10^{-10}~eV^2$ of the
$2\nu$ vacuum oscillation solution \cite{KP1}, 
however, solution B takes place for values of $\sin^22\theta_{12}$ which
are systematically smaller than the values of the same parameter
in the $2\nu$ vacuum oscillation solution. 

  The 0.862 MeV $^{7}$Be neutrinos take part in adiabatic MSW transitions
in the Sun, while the $^{8}$B neutrinos with $E \gtap 4~MeV$ undergo
nonadiabatic transitions. Both the $^{7}$Be and $^{8}$B neutrinos,
as well as the $\nu_{\mu}$ and/or $\nu_{\tau}$ into which a fraction of the $\nu_e$
has been converted by the MSW effect in the Sun, participate in vacuum oscillations
after leaving the Sun. These oscillations are modulated by the MSW probability
$\bar{P}^{(31)}_{2MSW}$. 
    
    With respect to the predictions in ref. \cite{BP95}, the signal in the 
Kamiokande detector and the contribution of the $^{8}$B 
neutrinos to the signals in the Cl--Ar
detector are smaller typically by factors of
$\sim (0.43 - 0.47)$ and $\sim (0.32 - 0.36)$. 
The $pp$ and the 0.862 MeV $^{7}$Be $\nu_e$ fluxes
are suppressed by factors of $\sim (0.65 - 0.90)$ and $\sim (0.11 - 0.27)$ for 
most of the values of the parameters from the allowed region. 
However, for $\sin^22\theta_{12} \sim 0.9$, for instance, one has 
$\bar{P}(\nu_e \rightarrow \nu_e ; t_E,t_0) \cong 0.55$ 
for the $pp$ neutrinos. Even in this case the 0.862 MeV $^{7}$Be $\nu_e$ flux 
is reduced by a factor of $\sim 0.3$, but the indicated possibility is rather
marginal.  

  The seasonal variations 
due to the vacuum oscillations 
of the signals in the Super-Kamiokande, SNO and ICARUS detectors are 
estimated to be smaller than the variations in the case of the $2\nu$ vacuum
oscillation solution \footnote{These variations were shown \cite{KP2} to be
not larger than 15\% for the $2\nu$ vacuum oscillation solution.},
except possibly in the small region
of the  $\Delta m^2_{21} - \sin^22\theta_{12}$ plane where 
$\Delta m^2_{21} \gtap 10^{-10}~eV^2$ and $\sin^22\theta_{12} \gtap 0.7$. 
The range of the predicted values
of the day-night asymmetry in these detectors is different from
the one expected for the $2\nu$ MSW solution.
The seasonal variation of the 0.862 MeV $^{7}$Be $\nu_e$ flux caused 
by the vacuum oscillations is expected to be considerably smaller than in the 
$2\nu$ case, while the day-night asymmetry is estimated to be somewhat 
smaller than the one predicted for the $2\nu$ MSW nonadiabatic solution.
The seasonal variation, nevertheless, may be observable.
Obviously, the experimental detection 
both of a deviation from the standard (geometrical) 6.68\% seasonal variation 
of the solar neutrino flux and of a nonzero day-night effect will be a proof 
that solar neutrinos take part in MSW transitions and vacuum oscillations.

\vskip 0.3cm   
{\bf Solution C and E}. 

The values of the parameters for solution C 
(Figs. 4a and 4b) form a rather large region in the 
$\Delta m^2_{31} - \sin^22\theta_{13}$ plane,
and a relatively small one in the 
$\Delta m^2_{21} - \sin^22\theta_{12}$ plane,
  The $\chi^2_{min}$ for this solution is larger than for solutions A and B:
$\chi^2_{min} \sim 1.5$ at 
($\Delta m^2_{21}$, $\sin^22\theta_{12}$, $\Delta m^2_{31}$, $\sin^22\theta_{13}$) 
$\cong$ ($1.2\times 10^{-10}~eV^2$, 0.78, $4.6\times 10^{-5}~eV^2$, 
$5.9\times 10^{-4}$) (the black square in Figs. 4). 

Solution E holds for a small region,
it and most of solution C are excluded by the recent experimental
spectrum data \cite{KAM}.

\vskip 0.3cm
{\bf Solution F}. 

This solution (FIG. 4a) was found in paper \cite{BQYAS}.  
For $\Delta m^2_{13}\sim (4~-8)\cdot10^{-6}$ eV$^2$,
the $^7Be$-flux can be suppressed by resonance conversion. Since
$~\sin^2 2\theta_{e\tau} \sim (3~-~10)\cdot 10^{-4}$, the pit is narrow
and suppression of the high energy part of the boron
neutrinos is rather weak. This flux can be suppressed by vacuum oscillations
if it is placed in the first minimum of oscillatory curve.  
For pp-neutrinos one gets then the averaged oscillation effect.
Thus we arrive at configuration with resonance conversion pit at small energies
and vacuum oscillation pit at high energies. 
\vskip 0.3cm
\leftline {\bf 4. Minimum $\chi^2$ value by including SuperKamiokande data}
\indent In model \cite{BP95}, using five experiment results (for SuperK, 
we use 201.6 live days data), we find the minimum $\chi^2$ of 
MSW+vacuum solution is 0.033, located at $(\Delta m^2_{21},
~\sin^2 2\theta_{12},~\Delta m^2_{31},~\sin^2 2\theta_{13})~ 
\approx~ (1.4\times 10^{-10}eV^2,~0.48,~9.9\times 10^{-6}eV^2,~4.0 
\times 10^{-3})$, which is in the B-down solution region.
Here we have four free parameters, thus it is one degree of freedom. 
\vskip 0.3cm
\leftline {\bf 5. Distortion of the boron neutrino spectrum and signals in}
{\bf SuperKamiokande and SNO}
\vskip 0.2truecm
An interplay of vacuum oscillations and MSW conversion can lead to
peculiar distortion of the boron neutrino energy spectrum. For
$\Delta m^2_{12} >
10^{-11}$ eV$^2$ there is modulation of the spectrum due to
vacuum oscillations.
We have studied a manifestation of such a distortion in the energy spectrum
of the recoil electrons in the SuperK (SuperKamiokande) and
SNO \cite{SNO} experiments.

Using energy resolution function for electrons \cite{KAM} we have found the
ratio
$R_e$ of expected (with conversion) number of events $S(E_{vis})$ to
predicted (without conversion) one
 for different values of oscillation parameters (FIG. 5):
\begin{equation}
\begin{array}{c}
$${\displaystyle S(E_{vis})~=~\int dE_e \cdot f(E_{vis};~E_e) \cdot
\int_{E_e-{m_e \over 2}} dE_{\nu} \cdot \Phi (E_{\nu})\cdot
~~~~~~~~~~~~~~~~~~~~~~~~~~~~~~~~~~~~~~~~~~~~}$$
\\ $$~~~~~~~~~~~~~~~~~~~~~~~~~~~~~~~~{\displaystyle \left[
P(\nu_e\rightarrow \nu_e) {d^2\sigma_{\nu_e} \over
{dE_e dE_{\nu}}}(E_e;~E_{\nu})~+~\nu_{\mu},\nu_{\tau}~{\rm contribution}
\right]},$$
\label{spectrum}
\end{array}
\end{equation}
where $E_e$ is the total energy of recoil electrons and the original
neutrino flux is $\Phi (E_{\nu})$. The lower limit of integration is 
first order approximation but the precise form is ${\displaystyle
{1 \over 2} \left ( E_e-m_e+\sqrt{E_e^2 - m_e^2} \right ) }$.	
The energy resolution function can be
written as \footnote{For ideal energy resolution, $f(E_{vis};E_e)$ goes to
$\delta$-function.}:
\begin{equation}
$${\displaystyle f(E_{vis};E_e)~=~
{1\over {\sqrt{2\pi}E_e\sigma}}\cdot exp\left (-\left ({E_{vis}-E_e \over
{\sqrt{2}E_e\sigma}}\right)^2\right)}.$$
\label{resolutionf}
\end{equation}
Experimental table of
$\sigma$ (for SuperK) is used in our calculation but
an approximate relation is
$\sigma \propto {1 \over \sqrt{E_e}}$.
We show also the $R_e$ measured by SuperK during 201.6 days \cite{SK2}.
As follows from FIG. 5, the integration over neutrino energy and on electron
energy weighted with resolution function leads to strong averaging out
of the oscillatory behaviour. Indeed, for present water Cherenkov experiment
like SuperK the energy resolution is typically $\sim~1.6$ MeV (at
$E_e=10$ MeV)
which is bigger or comparable with ``period'' (in the energy scale) of the
oscillatory curve. 

This hybrid solutions give very rich distortions of the recoil
spectrum. It not only has the pure 2-$\nu$ solution's distortions but also
give other peculiar distortions.

\begin{figure}[H]
\mbox{\epsfig{figure=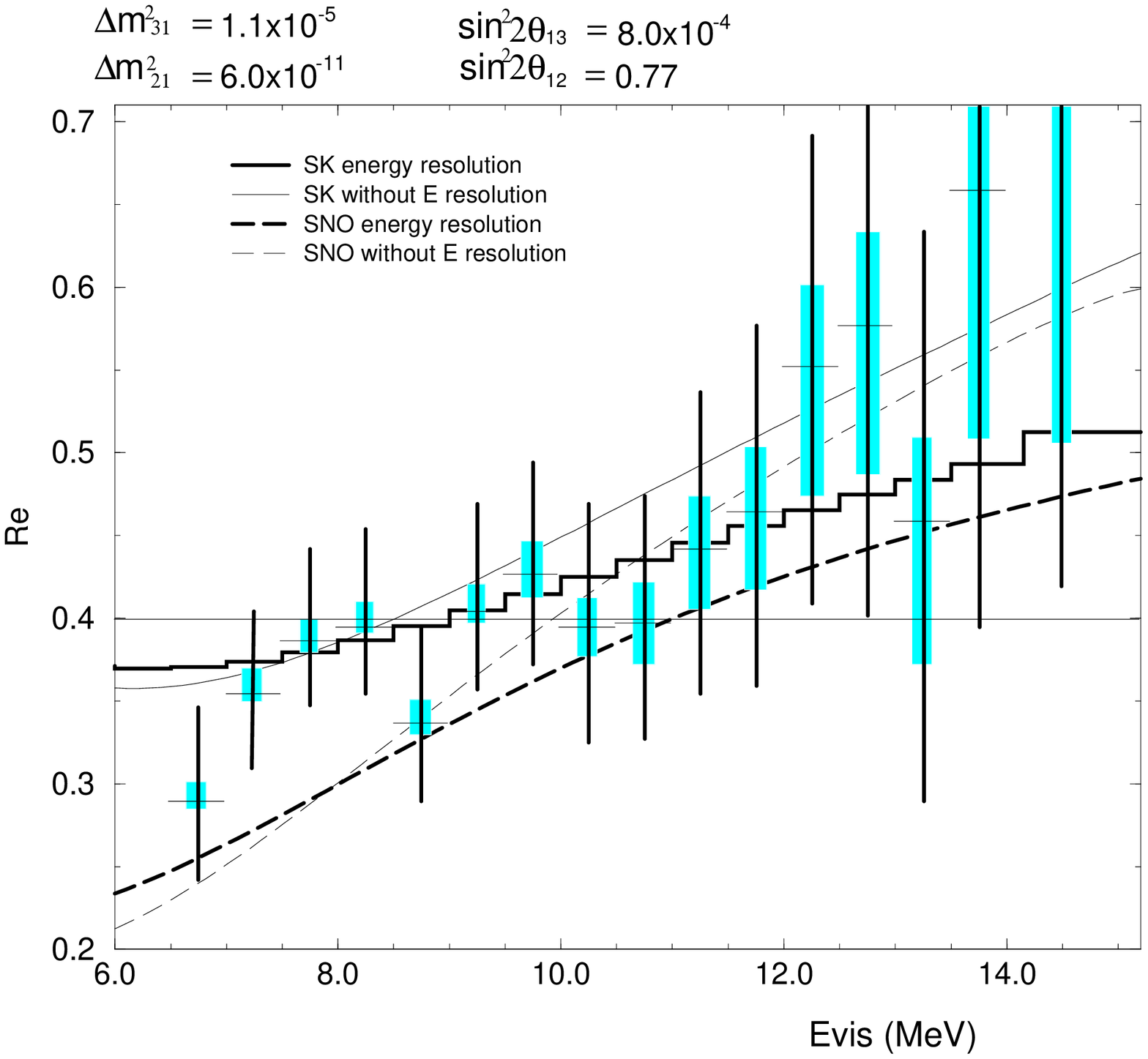,height=7.3cm,angle=-90}}
\mbox{\epsfig{figure=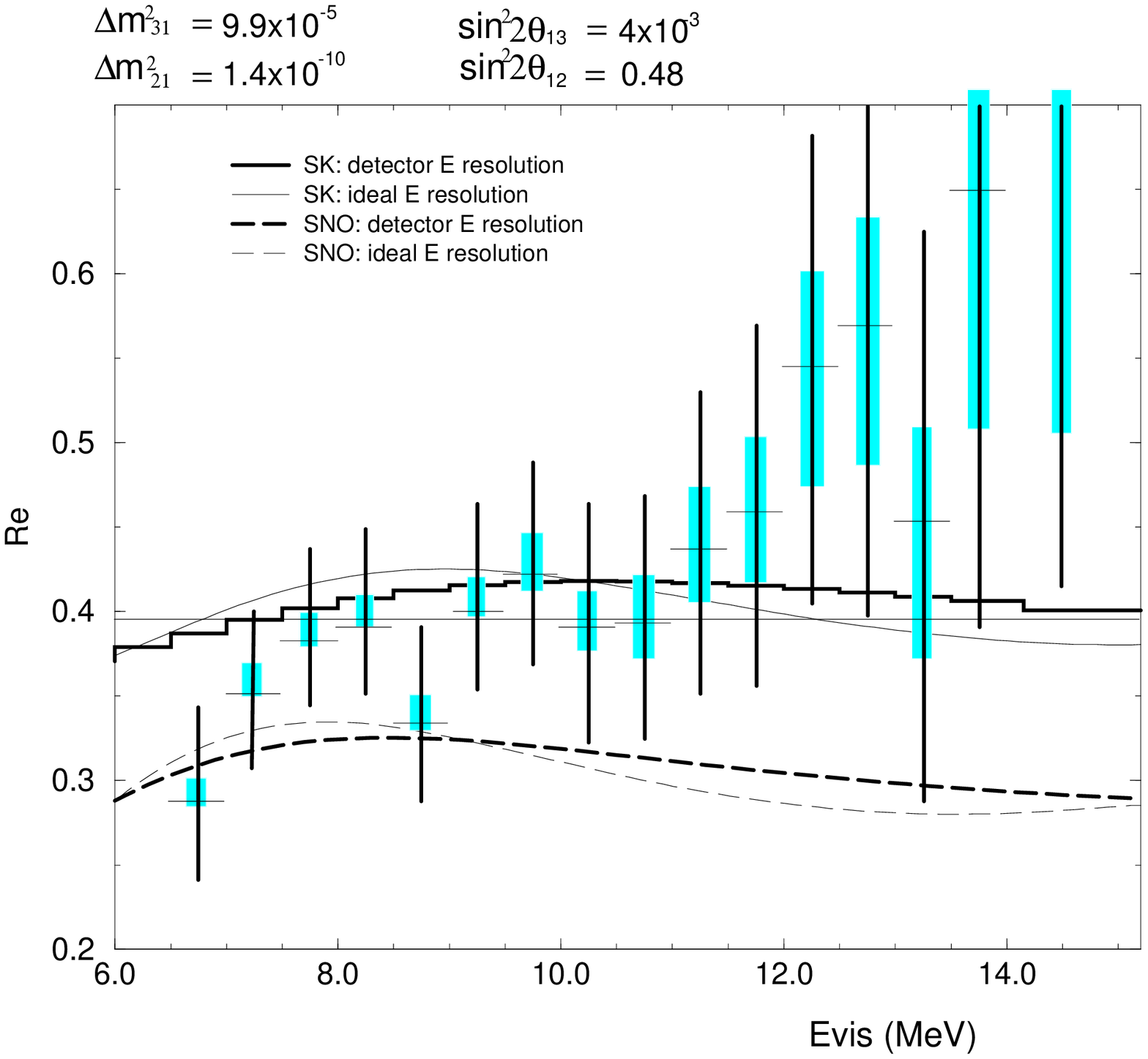,height=7.3cm,angle=-90}}
\vglue 0.2cm
\caption[]{ The expected spectrum deformations of the recoil electrons
in SuperK and SNO. Boxes together with solid lines are SuperK data. } 
\end{figure}
 
Obviously some of distortions like in region E
are already excluded by 201.6 days data. The smoothing
effect is weaker in SNO experiment: The intergration over neutrino energy gives
smaller averaging and energy resolution is slightly better. 

 \vglue 0.4cm
\leftline{\bf 6. Other case of MSW+vacuum solution.}
  Besides three active neutrinos, other cases can provide mixed solutions
also. For instance, the mass scheme \cite{QYAS1} below can give
2 active + 1 sterile neutrinos to solve solar neutrino problem (FIG. 6).

\begin{figure}[H]
\mbox{\epsfig{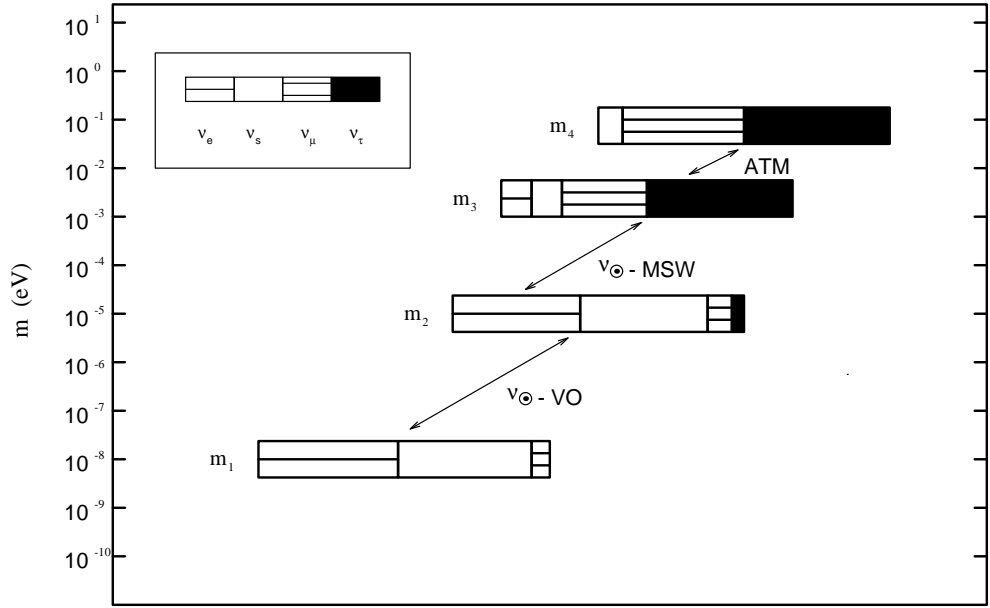}}
\caption[]{ Qualitative pattern of the neutrino masses and
mixing. Boxes correspond to different mass eigenstates. The sizes of
different regions in the boxes determine flavors ($|U_{if}|^2$) of
given eigenstates. Arrows connect the eigenstates involved in 
ATM - atmospheric oscillation, $\nu_\odot$-MSW - MSW conversion 
inside the Sun, $\nu_\odot$-VO - vacuum oscillation between Sun and
Earth. }
\end{figure}

\indent Suppose there are four neutrinos in nature, three independent 
\dm have magnitudes $10^{-11}eV^2$, $10^{-5}eV^2$ and $10^{-2}eV^2$. 
The biggest
\dm gives solution to atmospheric and LSND anomalies via 
$\nu_{\mu}~\leftrightarrow
~ \nu_{\tau}$ vacuum oscillation. The middle \dm provide MSW transition 
($\nu_e ~ \leftrightarrow ~ mixed~ state~ of~ \nu_{\mu},~\nu_{\tau}$)  
inside the sun.
And the smallest \dm gives vacuum oscillation $\nu_e ~ \leftrightarrow 
~ \nu'_s$ which is $\nu_{sterile}$ mixed slightly with $\nu_{\mu}$ during the 
travel from the surface of the Sun to the surface of the Earth.

  \vglue 0.4cm
\leftline{\bf 7. Conclusions.} 
\vskip 0.3cm
  
    A general feature of the MSW + VO 
solutions studied by us is that
the $pp~~\nu_e$ flux is suppressed (albeit not strongly - 
by a factor not smaller than 0.5) 
primarily due to the 
vacuum oscillations of the $\nu_e$, the suppression of the 
0.862 MeV $^{7}$Be $\nu_e$ flux 
is caused either by the vacuum oscillations or by the combined effect of the 
MSW transitions and the vacuum oscillations, 
while the $^{8}$B $\nu_e$ flux is suppressed either due to the
MSW transitions only or by the interplay of the MSW transitions 
in the Sun and the oscillations in vacuum on the way to the Earth. 
The solutions differ in the way the $pp$, $^{7}$Be and 
the $^{8}$B neutrinos are affected by the $\nu_e$
MSW transitions and/or the oscillations in vacuum.

  For all MSW + VO solutions we have 
considered, the $^{8}$B $\nu_e$ spectrum is predicted to be rather strongly 
deformed. 
The SuperK data 
on the shape of the $^{8}$B neutrino spectrum can be used to further constrain the
solutions we have found. Such an analysis can exclude solution E, almost all
of C and some part of A,B,D,F solutions.

   For the MSW + VO solutions considered by us
the day-night asymmetry in the signals of the detectors sensitive only to
$^{8}$B or $^{7}$Be neutrinos are estimated to be rather small, not exceeding
a few percent. The seasonal variation effect caused by the vacuum oscillations
can be observable for $^{7}$Be neutrinos and, for certain 
relatively small regions of the allowed values of the parameters, 
can also be observable for the $^{8}$B or for 
the $pp$ neutrinos if the $pp$  neutrino flux 
is measured with detectors like HELLAZ or HERON.  
 
  The global minimum $\chi^2$ of this hybrid solution is very small. 
The interplay of the resonance conversion and vacuum oscillations leads to
additional peculiar distortion of the neutrino energy spectrum. 
In the SuperKamiokande experiment, and (to a slightly weaker extend)
in the SNO, the integrations over the neutrino energy and 
finite energy resolution
result in strong smoothing of oscillatory distortion of the electron energy
spectrum.

  Finally, the MSW transitions + vacuum oscillation solutions can also
be obtained from
the neutrino mass and mixing structure in the case
of two active plus one sterile neutrinos.

\vglue 0.4cm
\leftline{\bf Acknowledgements.} Most of the work are from paper 
\cite{QLSP} which written with S.T. Petcov and \cite{BQYAS} with
A. Yu. Smirnov and K.S. Babu.
The author would like to thank J. Bahcall for useful information and
W. Hampel for warm hospitality in this conference. 
In particular, I appreciatively thank to A.Yu. Smirnov for all kinds of help.
The work was supported in part 
by the EEC grant ERBFMRXCT960090.

\end{document}